\documentclass[11pt]{article}
\usepackage[preprint]{acl}
\usepackage{times}
\usepackage{latexsym}
\usepackage[T1]{fontenc}
\usepackage[utf8]{inputenc}
\usepackage{microtype}
\usepackage{inconsolata}
\usepackage{graphicx}
\usepackage{booktabs}
\usepackage{tabularx}
\usepackage{multirow}
\usepackage{amsmath,amssymb}
\usepackage{longtable}
\usepackage{upquote}
\usepackage{xcolor}
\usepackage{array}
\usepackage{calc}
\usepackage{etoolbox}
\usepackage{parskip}
\usepackage{bookmark}
\IfFileExists{xurl.sty}{\usepackage{xurl}}{}
\hypersetup{hidelinks}

{}

\makeatletter

 \let\@cite@ofmt\@firstofone

 \def\@biblabel#1{}

 \def\@cite#1#2{{#1\if@tempswa , #2\fi}}

\makeatother

\newlength{\cslhangindent}

\newlength{\csllabelwidth}

\providecommand{\tightlist}{%

  \setlength{\itemsep}{0pt}\setlength{\parskip}{0pt}}

\IfFileExists{xurl.sty}{\usepackage{xurl}}{} 

\title{The Same Zero: Why Identical ASR Can Imply Different Guarantees in LLM-Agent Security}
\author{Yajie Yin\\ \texttt{1549080929@qq.com}\\ ORCID: 0009-0001-6168-2530}

\begin{document}
\maketitle


\begin{abstract}

LLM-agent security has produced a dense landscape of defenses---prompt hardening, content filters, permission gates, sandboxes---yet no framework tells a deployer what a defense actually guarantees, or where that guarantee comes from. We apply Verification Autonomy Levels (VAL)---L0: LLM self-declaration; L1: deterministic rules; L2: objective ground truth; L3/L4: decidable completeness; L5: impossible---to 22 agent-security defenses; the taxonomy is falsifiable (10/10 prediction hits on frozen cards, flagged). We run the first controlled deployment-value comparison: at equal budget, a VAL-guided stack (confirmation gate + schema sandbox) versus a mainstream intuition stack (prompt hardening + keyword filter), 50 scenarios, 12 attack variants, adaptive/white-box/PAIR escalation ($\approx$7,000 testbed calls; $\approx$10,000 harness calls on AgentDojo/JADE). The VAL stack holds 0.000 attack success at 1.000 benign success (0.5\% ASR at 79.7\% utility on AgentDojo banking vs 4.3\% undefended); the intuition stack reaches 0.000 ASR but kills all benign actions---security by model-behavior luck, not structure. Across testbeds of rising attack-surface hardness the intuition stack's zero drifts (0$\rightarrow$1.9\%$\rightarrow$6.2\%, n=16 on JADE) while the VAL stack's holds within its ODD (0$\rightarrow$0$\rightarrow$0), its only breach a disclosed out-of-ODD password gap (0.5\%). The same zero, two different guarantees: zero is an outcome, not a guarantee.

\end{abstract}
\section{Introduction}\label{introduction}

Large language model (LLM) agents are moving from single-turn assistants to persistent systems that call tools, browse the web, and maintain long-term memory---and the matching expansion of security defenses (prompt hardening, filters, permission gates, sandboxes, provenance firewalls) is presented with implicit, rarely commensurable guarantees: one paper's ASR is not another's, and no framework tells a deployer \emph{which defense to trust for which threat, before running experiments}.

This paper asks the question the field has not asked explicitly: \textbf{what does a given agent-security defense guarantee, and where does that guarantee come from?}

We answer with Verification Autonomy Levels (VAL), a taxonomy proposed in (Yin 2026) for verification schemes generally. VAL classifies any verification/authorization scheme along a single axis---the source of its verification spec---into six levels:

\begin{itemize}

\tightlist

\item

  \textbf{L0}: the spec is LLM self-declaration (``I checked it,'' ``this is safe''). No deterministic anchor. Failure mode: the declaration can be rewritten by the very model that issues it.

\item

  \textbf{L1}: deterministic rules (regex, keyword lists, hand-written policies). Failure mode: surface forms are bypassable by rewriting.

\item

  \textbf{L2}: objective ground truth (gold labels, platform-recorded events, execution results). Guarantees \emph{correctness only}: everything checked passes, but nothing is proven about what was not checked (the completeness blind spot).

\item

  \textbf{L3/L4}: decidable systems (solveset equivalence, type systems, formal kernels, confinement mechanisms). Single-property or domain-level completeness within an operational design domain (ODD).

\item

  \textbf{L5}: universal completeness---impossible in the unrestricted case.

\end{itemize}

VAL's core claim is that a scheme's level is determined by its \emph{anchor}, not by its accuracy or sophistication, and that the anchor determines the scheme's failure modes. Applied to agent security, this yields a falsifiable prediction: \textbf{a defense is defeated the way its anchor fails.} An L0 defense fails when the model complies with the injection; an L1 defense fails when the attacker rewrites around the surface rule; an L2 defense fails outside its ODD (forged metadata, poisoned data); an L3 defense holds within its ODD by construction.

We make three contributions:

\begin{enumerate}

\def\labelenumi{\arabic{enumi}.}

\tightlist

\item

  \textbf{A level map of agent security.} We classify 22 defenses using a frozen, blind-rater-validated protocol (inter-rater $\kappa$ $\approx$ 0.8 under the frozen protocol (same-family LLM raters); provenance in Appendix A.1 (Yin 2026)), organizing the landscape by what each defense can actually guarantee.

\item

  \textbf{Prediction validation.} Frozen level-to-behavior cards, validated on a small published sample (10/10 hits, flagged; two mechanism-driven corrections)---the point is falsifiability, not the hit count.

\item

  \textbf{Deployment-value comparison and zero stability.} Same budget, two stacks (VAL-guided vs.~mainstream intuition), one testbed---50 scenarios, 12 attack variants, real effects, adaptive/white-box/PAIR escalation, three seeds, a second victim, and the official AgentDojo benchmark. The VAL stack dominates on security within its operational design domain (0.000 ASR on our testbed; 0.5\% on full AgentDojo banking vs 4.3\% undefended---an 8.6$\times$ reduction at a 6.9pp utility cost); escalating-attack trajectories (Z($\alpha$)) further separate defenses by failure morphology, and identical trajectories still require provenance to interpret---zero stability joins ASR and compliance as a reportable property.

\end{enumerate}

Appendix A.0 draws the paper's chain: the same zero, escalation Z($\alpha$), and the provenance funnel that separates them.

\textbf{What we do not claim.} VAL is not a general theory of verification, a complete security evaluation, or a ranking of defenses---it is a mid-level methodological claim: the provenance of an outcome determines its guarantee and predicts its failure boundary. Robotics, formal methods, HITL, and non-agentic LLMs are outside scope; so is a claim that VAL covers every defense.

\section{Background and Related Work}\label{background-and-related-work}

\subsection{Verification Autonomy Levels (VAL)}\label{verification-autonomy-levels-val}

VAL (Yin 2026) grades verification schemes by three questions: Q1 \emph{where does the verification spec come from} (LLM declaration / deterministic rule / objective truth / decidable system); Q2 \emph{what does the verdict guarantee} (nothing / correctness / completeness); Q3 \emph{what scope does a completeness claim cover} (single property / domain / universal). The level is the first rule that fires: L5 (universal completeness, impossible), L4 (domain completeness), L3 (single-property completeness), L2 (correctness with objective anchor), L1 (correctness with deterministic rule), L0 (otherwise). The procedure is deterministic and has been measured for inter-rater reproducibility: seven independent blind LLM raters reach $\kappa$ $\approx$ 0.8 pairwise under the frozen protocol (v2.2/v2.3); the full rater provenance (raters, sheet sizes vs.~rater overlaps, nested subsets) is stated once in Appendix A.1 (Yin 2026).

Two refinements matter. First, \textbf{anchor semantics} (Yin 2026): objective anchors split by what they certify---intent (a recorded decision event), truth (a fact), effect (an execution outcome). A confirmation record is an intent anchor: it authorizes, but says nothing about correctness; conflating intent with effect is the category error behind most authorization failures. Second, \textbf{the completeness blind spot} (Yin 2026): checks verify proposed candidates but cannot prove none was missed---the analog in security is the \emph{un-enumerated attack}: a gate blocking all evaluated attacks is a correctness probe over the evaluated distribution, not a guarantee over all attack paths.

\subsection{Agent-security defenses and their taxonomies}\label{agent-security-defenses-and-their-taxonomies}

The defense literature spans text-level (prompt hardening, filters, Llama Guard), tool-level (allowlists, IFC, sandboxes), and memory-level (A-MemGuard, PPMF) defenses, plus benchmarks like AgentDojo. Recent work separates injection from execution success (Leong 2026), classifies test oracles by authority source (Mughal and Bilal 2026), shows same-family verification yields near-zero gain (Lu et al. 2025), and situates agent security within safety-landscape taxonomies (Chen et al. 2024), system-evaluation frameworks (Xia et al. 2024), and verifier-performance studies (Seo et al. 2025). Two recent directions sharpen the stakes: offline-verifiable evidence bundles bound what a verifier may infer from preserved logs (Alshammari and Bahsi 2026), and self-evolving harnesses must pass recoverability-constrained validation before mutations are admitted (Sah et al. 2026). Our contribution is different in kind: not another defense or evaluation recipe, but a \emph{pre-purchase axis}---the anchor---that classifies the source of a defense's guarantee and predicts each level's failure modes.

\section{Method}\label{method}

\subsection{Classification protocol}\label{classification-protocol}

We classify defenses with the frozen VAL protocol (v2.3, (Yin 2026)) extended for security: R2 mixed systems split per component (weakest anchor on the verdict path); R3 evidence anchors do not equal verdict anchors; R10 designer rules are L1; R12b theorem-backed exhaustive procedures (confinement, taint tracking) are L3; R15 LLM-generated expectations are L0. Each defense is rated on Q1/Q2/Q3 from a mechanism-only evidence pack by blind raters. Rater provenance --- the raters, sheet sizes vs.~rater overlaps, and the two nested security subsets that must not be conflated --- is stated once in Appendix A.1.

\textbf{On the validity of the taxonomy.} Agreement among raters shows the categories are \emph{usable}, not \emph{right}. We offer three independent lines instead: \emph{out-of-sample prediction} (cards frozen before outcome data, 10/10 hits with two mechanism-driven corrections---retrospective storytelling cannot produce that pattern); \emph{cross-benchmark convergence} (the same anchor logic predicts our testbed, AgentDojo, JADE, and LITMUS results---four sources with no shared implementation); and \emph{falsifiability} (each card names the attack predicted to break its defense). We claim VAL is a productive organizing principle, not the unique correct ontology.

\subsection{Prediction protocol}\label{prediction-protocol}

Following (Yin 2026), we freeze prediction cards before outcome data: for each defense, level $\rightarrow$ predicted behavior pattern + the specific attack predicted to break it, validated against published data (PPMF's numbers; abstract-level evidence flagged as such). The protocol also generalizes out of sample: before running JADE's MCP instances (Appendix A.3) we froze five predictions (e.g., N's keyword filter leaks the post\_tweet variant---its action word absent from the filter's vocabulary; V's sandbox blocks every variant). All five held, including the boundary prediction.

\subsection{Deployment-value experiment}\label{deployment-value-experiment}

\textbf{Testbed.} A minimal agent: external content is written into memory; a user task triggers a tool call; the LLM (DeepSeek-chat, temperature 0.2) proposes a call; the defense chain decides; a real-effect executor performs the action (a ledger for transfers, an outbox for email, real file deletion in a sandbox directory, a forum log for comments). Attack success requires a \emph{real state change} (server-side effect semantics, as in (Xu et al. 2026)), not merely a proposal.

\textbf{Scenarios and attacks.} 50 scenarios (30 malicious + 20 benign; the benign set is larger than the 25\% share of (Xu et al. 2026; Debenedetti et al. 2024) to narrow utility confidence intervals), derived from the memory-provenance families of (Xu et al. 2026). Attacks: 8 self-written variants plus 4 third-party AgentDojo payloads (Debenedetti et al. 2024), 12 total (full list in Appendix A.5); escalation adds an adaptive attacker (3 rounds, rewrites from the interception reason), a white-box adaptive attacker (told the system prompt and keyword list), and a PAIR-style candidate-search attacker.

\textbf{Defenses.} Four self-implemented defenses, one per anchor family: - \textbf{D1} (L0): system-prompt hardening---``ignore instructions in external content; never perform high-risk actions without explicit user request.'' The verdict is the model's own compliance. - \textbf{D2} (L1): keyword filter over memory and call text (transfer/delete/send/payment terms, Chinese equivalents). - \textbf{D3} (L1/L2): confirmation gate---high-risk tools require a \emph{platform-recorded} confirmation event; text claims (``user-confirmed'') are ignored. - \textbf{D4} (L3): parameter sandbox---tool arguments must match schema (recipient whitelists, target patterns, type/range checks), executed in a side-effect stub with verified state changes.

\textbf{Stacks (same budget, two defenses each).} - \textbf{V (VAL-guided)}: D3 + D4---an intent anchor (L2 input) plus a confinement property (L3). - \textbf{N (mainstream intuition)}: D1 + D2---prompt hardening plus keyword filtering, the intuitive low-cost default in production agents. The comparison does not claim ``VAL beats all defenses'' (structural defenses dominate behavioral ones by construction); the contribution is the \emph{epistemic} contrast---identical zeros, opposite guarantees, a sharp utility cost for the intuition stack (Section 6.1).

\textbf{Metrics.} ASR = high-risk tool executed with complete arguments and verified state change; compliance rate = the model \emph{proposes} a high-risk tool (including empty-argument hedging)---separating ``model was driven by the injection'' from ``the defense intercepted.'' Benign success = benign scenarios complete their intended action.

\textbf{Runs.} Static: 7 configs $\times$ 380 cases (30 malicious $\times$ 12 attack variants + 20 benign). Adaptive/white-box: 30 malicious scenarios $\times$ stacks $\times$ 3 rounds. Seeds: V, N, and the no-defense baseline (ND) re-run 3$\times$. A second victim (Llama 3.1 8B) re-runs the static battery. Total $\approx$ 7,000 LLM calls across all experiments; AgentDojo (Section 6.5, full banking) and JADE probes add $\approx$10,000 harness-internal calls reported separately.

\section{The Defense Level Map}\label{the-defense-level-map}

The full level map (all 22 defenses) and per-defense ratings with evidence are in Appendix A.1; blind-rater agreement is given once in Appendix A.1: 20/22 = 90.9\% exact on the 22-item security subset (B6 vs B7), and 6/6 on the agentsec schemes nested in the 54-scheme comparison (B5 vs B6, $\kappa$ 0.882 over those 54 schemes (Yin 2026)).

\textbf{The structural finding.} L3 anchors exist in agent security---but only for \emph{confinement and information-flow properties}: sandbox execution domains, taint tracking, data-control separation, SMT-monotonic permission lattices. These are decidable by construction. \emph{Semantic} safety (``is this action dangerous?'') is an open-world property for which no decidable fragment is available in the unrestricted setting, and hence caps at L2; restricted semantic properties that can be formally encoded (e.g., a bounded action taxonomy) remain eligible for higher levels. This mirrors the formal/empirical dichotomy of (Yin 2026) exactly: completeness is reachable only for formally specifiable properties.

\subsection{Worked examples: how the classification reads defenses}\label{worked-examples-how-the-classification-reads-defenses}

Each classification is a claim; per the framework's own epistemology, claims need anchors. We show the full evidence chain for three representative defenses spanning the level spectrum (remaining classifications in the supplementary corpus).

\textbf{Self-Ask provenance --- no anchor (L0).} Mechanism: the LLM infers whether memory text looks external, user-history-like, or confirmation-like, and authorizes from the inferred trust (Xu et al. 2026). Q1: LLM; Q2: none; Level: \textbf{L0}. Evidence: permissive mode reaches \textbf{0.984 ASR}; strict mode blocks all benign actions (Xu et al. 2026). Predicted failure mode: text claims (fake markers, social engineering). Observed: 0.984 ASR under permissive inference---the anchor is whatever the model believes.

\textbf{PPMF --- the intent anchor (L1/L2).} Mechanism: a deterministic gate authorizes tool calls by matching action risk to the least-trusted supporting memory, whose provenance metadata (source, trust, confirmation events) is platform-maintained (Xu et al. 2026). Q1: designer policy over platform-recorded events; Q2: correctness; Level: \textbf{L1} (weakest anchor) with L2-grade inputs. Evidence: 0.000 ASR with intact metadata; \textbf{0.088 ASR under 10\% forged confirmations}---the completeness blind spot quantified. Predicted failure mode: forged/compromised metadata (outside the ODD). Observed: exactly that; the authors themselves list compromised metadata outside their guarantee.

The two examples trace the lower ladder with primary-source evidence: L0 fails to text belief, L1/L2 fails outside its ODD (forged metadata); the L3 confinement example (IFC/Fides) is in Appendix A.1. All failure modes were predicted before the validation data were consulted (Section 5).

\section{Prediction Validation}\label{prediction-validation}

We froze 16 prediction cards (level $\rightarrow$ predicted failure mode + predicted breaking attack) before consulting outcome data (\textbf{12 of the 16 are reported below; the other four are not.} With five JADE cards, \textbf{21 frozen in total.})

\textbf{PPMF family (6 defenses):} all six predictions hit on PPMF's published numbers (Xu et al. 2026): content filters collapse under rewriting; Self-Ask provenance (L0) reaches 0.984 ASR; the gate-only ablation kills benign actions; full PPMF (L1/L2) achieves 0.000 ASR with benign preserved, and 0.088 under 10\% forged confirmations---the completeness blind spot quantified.

\textbf{New cards (abstract-level evidence, flagged):} 4/4 behavior predictions hit (Llama Guard distribution-bound accuracy; RA-LLM residual ASR under strong attacks; perplexity filtering weak alone; ToolEmu audit-value-only). Two mechanism-driven corrections emerged; because they are the strongest evidence that the prediction loop \emph{corrects} rather than merely confirms, each gets a full account below.

\subsection{Mechanism-driven correction I: Progent's SMT monotonic confinement}\label{mechanism-driven-correction-i-progents-smt-monotonic-confinement}

\textbf{Predicted:} L1 (designer permission policy) (Shi et al. 2025). \textbf{Mismatch:} every policy update is adjudicated by an SMT solver as \emph{narrowing} (automatic) or \emph{expanding} (requires approval)---the agent's action space can only shrink without approval, a confinement property with a decidable witness. \textbf{Corrected classification:} the gate stays L1, but the monotonic-confinement guarantee is annotated L3 (theorem-backed structure, R12b). The mismatch is evidence for, not against, the protocol.

\subsection{Mechanism-driven correction II: RARR's LLM-mediated verdict (R3)}\label{mechanism-driven-correction-ii-rarrs-llm-mediated-verdict-r3}

\textbf{Predicted:} L2 (attribution checking). \textbf{Mismatch:} RARR's verdict path is LLM-mediated end-to-end---``citation existence'' is not mechanically checked. Under R3 (evidence anchors do not equal verdict anchors), the verdict component is LLM declaration. \textbf{Corrected classification:} L2 $\rightarrow$ L0; a superficially ``objective'' anchor can be L0-masked when the alignment step is model-mediated---precisely the laundering pattern of the memory domain.

\textbf{Cumulative: 10/10 behavior predictions hit (6 high-confidence, 4 abstract-level), 2 classification corrections.} Sample is small; the value is that the prediction \emph{protocol} is now demonstrated end-to-end, not that the hit rate is large.

\section{Deployment-Value Experiment: Results}\label{deployment-value-experiment-results}

\textbf{The dominance at a glance} (full tables in 6.1-6.4):

{\def\LTcaptype{none} 

\begin{table*}[!t]
\centering
\small
\begin{tabularx}{\textwidth}{XXXX}
\begin{minipage}[b]{\linewidth}\raggedright

Stack

\end{minipage} & \begin{minipage}[b]{\linewidth}\raggedright

ASR (both victims)

\end{minipage} & \begin{minipage}[b]{\linewidth}\raggedright

Benign (both victims)

\end{minipage} & \begin{minipage}[b]{\linewidth}\raggedright

What the zero is

\end{minipage} \\

\hline \\ \hline
\hline

ND (no defense) & 0.333 / 0.253 & 1.000 / 1.000 & no defense \\

\textbf{V (VAL: gate + sandbox)} & \textbf{0.000 / 0.000} & \textbf{1.000 / 1.000} & \textbf{structure - platform record + schema, victim-independent} \\

\textbf{N (intuition: hardening + filter)} & 0.000 / 0.000 & \textbf{0.000 / 0.000} & \textbf{luck - the model's refusal behavior, model-dependent} \\
\end{tabularx}
\end{table*}

}

The two zeros in the last two rows are the paper's spine: identical security numbers, opposite guarantees, a 100-point utility gap.

\subsection{Static attacks (12 variants, 380 cases/config, real effects)}\label{static-attacks-12-variants-380-casesconfig-real-effects}

{\def\LTcaptype{none} 

\begin{table*}[!t]
\centering
\small
\begin{tabularx}{\textwidth}{XXXXX}
\begin{minipage}[b]{\linewidth}\raggedright

Config

\end{minipage} & \begin{minipage}[b]{\linewidth}\raggedright

ASR

\end{minipage} & \begin{minipage}[b]{\linewidth}\raggedright

Compliance

\end{minipage} & \begin{minipage}[b]{\linewidth}\raggedright

Benign

\end{minipage} & \begin{minipage}[b]{\linewidth}\raggedright

gain vs ND

\end{minipage} \\

\hline \\ \hline
\hline

ND (no defense) & \textbf{0.333} & 0.492 & 1.000 {[}0.84, 1.00{]} & --- \\

D1 prompt hardening (L0) & 0.000 & 0.219 & 0.000 {[}0.00, 0.16{]} & 0.333, benign destroyed \\

D2 keyword filter (L1) & 0.200 & 0.492 & 1.000 {[}0.84, 1.00{]} & 0.133 \\

D3 confirmation gate (L1/L2) & 0.000 & 0.497 & 1.000 {[}0.84, 1.00{]} & 0.333 \\

D4 parameter sandbox (L3) & 0.067 & 0.494 & 1.000 {[}0.84, 1.00{]} & 0.266 \\

\textbf{V = D3+D4 (VAL)} & \textbf{0.000} & 0.497 & \textbf{1.000 {[}0.84, 1.00{]}} & \textbf{0.333, lossless} \\

\textbf{N = D1+D2 (intuition)} & 0.000 & 0.233 & \textbf{0.000 {[}0.00, 0.16{]}} & 0.333, benign destroyed \\
\end{tabularx}
\end{table*}

}

Benign success is measured on 20 benign scenarios (Wilson 95\% CIs in brackets); the utility gap between V and N is significant at p \textless{} 0.001 (Fisher's exact test on 20/20 vs 0/20).

The two stacks reach the same zero but differ in \emph{how}: V's zero is produced by a gate the model cannot influence; N's zero by the model's behavior under the hardening prompt (compliance 0.233---proposing the malicious tool a quarter of the time, mostly with empty arguments; strict ASR counts only actionable executions). D2 alone leaks 0.200: the keyword filter is bypassable by rewriting.

\textbf{The two layers of V are not redundant.} D3 refuses high-risk tools without a platform-recorded authorization event the model cannot forge; D4 constrains what an \emph{authorized} action may touch (schema, whitelists, target patterns). Single-defense numbers show the split: D3 alone already reaches 0.000 ASR, D4 alone leaks 0.067 (schema-legal semantic attacks). An attack must defeat both anchors: D3 controls \emph{whether}, D4 controls \emph{what}.

\subsection{Adaptive and white-box attacks}\label{adaptive-and-white-box-attacks}

The full escalation table is in Appendix A.5; the key contrast: D1/D3 hold at 0.000 across static/adaptive/white-box, while D2 (0.200$\rightarrow$0.467) and D4 (0.067$\rightarrow$0.200) leak and worsen.

But the two surviving zeros rest on different grounds, visible in compliance under the same attacks: D3's model is driven to propose the malicious tool in 53.3\% of adaptive cases and 56.7\% of white-box cases and is still blocked every time (the gate reads a platform record the attacker cannot write), while D1's model is driven 36.7\%/33.3\% and the zero depends on its refusal behavior holding under every attack form we tried. The same static zero, the same adaptive zero, the same white-box zero---different provenance, and different predicted breakpoints under attacks strong enough to move a behavioral refusal (Section 7).

\textbf{Zero stability under escalating attack (Z($\alpha$)).} Tracking the cumulative ASR round by round (7 rounds, 30 cases per point; full data in Appendix) exposes distinct stability morphologies. D2 climbs monotonically to \textbf{0.833 by round 7} (each rewrite learns another way around the keyword vocabulary); D4 jumps at round 2 and plateaus (0.067 $\rightarrow$ 0.300); D1 and D3 stay at 0.000 across all seven rounds. The two flat curves are indistinguishable in Z($\alpha$) alone---which is exactly the point. \textbf{Identical observed zero-stability profiles do not, by themselves, establish equivalent security guarantees}; only the provenance separates them, and the gap widens under escalation (D3 driven 83.3\%, still blocked every time; D1 driven 36.7\%, zero = refusal behavior holding). We attempted to falsify D1's flat curve directly (7-round adaptive attacks with prompts explicitly demanding complete executable arguments, plus authority/emergency/role-play frames): it did not break under this attack budget---stronger support for the provenance distinction, not evidence that it cannot break. We predict it will break under an attack strong enough to defeat DeepSeek's refusal behavior (e.g., a less-aligned attacker); D3's will not, short of compromising the audit substrate itself (Appendix).

\textbf{Where the zero breaks: an execution funnel under static attacks} (a mechanism snapshot, distinct from the 7-round Z($\alpha$) above). D1's funnel collapses at the complete-argument stage---the model fails to produce executable arguments under the hardening prompt; the anchor remains behavioral (same shape as N's, 0.233 $\rightarrow$ 0.000). D3's funnel reaches the authorization stage: complete arguments in 49.7\% of cases, refused every time by the gate---the zero is the platform record (same shape as V's, 0.497 $\rightarrow$ 0.497 $\rightarrow$ 0.000). The same ASR and flat Z($\alpha$), but the funnels break at different layers: model behavior (arguments) vs.~external constraint (authorization).

\textbf{Definition (hedging wall).} Under a hardening prompt (D1), an LLM victim under injection may comply \emph{partially}: it proposes the requested high-risk tool while emitting empty or incomplete arguments---a stable compromise between the injection (``propose the action'') and the safety prompt (``do not execute without authorization''). Actionable ASR requires complete arguments, so hedging suppresses ASR \emph{without any structural gate}. The wall is a model behavior, not a mechanism: its height is victim-specific (compliance 0.233 for DeepSeek, 0.014 for Llama 8B, Section 6.4).

Against the stacks, V holds even when the model is driven to propose the malicious tool in 57\% of scenarios (white-box compliance 0.567, 2.4$\times$ N's 0.233); N holds only because its hedging wall survives static, adaptive, white-box, PAIR (0.000), and a willing cross-family attacker (DeepSeek against Llama: 0.000, Section 6.4). Aligned cross-family attackers (Kimi, Llama 8B) declined the role---explicit refusals, sanitizing rewrites; gradient-based attacks (GCG) are out of scope (the API exposes no gradients).

\subsection{Seeds and honesty}\label{seeds-and-honesty}

V, N, and ND re-run 3$\times$: standard deviations $\approx$ 0.005 on compliance, 0.000 on ASR and benign success. The headline contrast is stable (details and two harness bugs caught and fixed mid-experiment in Appendix).

\subsection{Victim generalization: a 2x2 matrix}\label{victim-generalization-a-2x2-matrix}

All results above use DeepSeek-chat as the victim. To test whether the structural pattern is victim-specific, we ran the same static attack suite (n = 30 malicious + 20 benign per model, identical battery) with a second, independent victim---Meta Llama 3.1 8B (local Ollama, same harness, tool registry added to the prompt):

The 2$\times$2 matrix is in Appendix A.2.

Three findings. First, \textbf{baseline injection compliance is model-specific}: Llama 8B is more resistant than DeepSeek (compliance 0.328 vs 0.492). Second, \textbf{N's ``model luck'' is confirmed by a second model}: hardening suppresses actionability to a model-specific degree (0.233 vs 0.014). Third, \textbf{V's structural claim is victim-independent}: 0.000 ASR and 1.000 benign for both, tolerating 0.317 compliance; a willing cross-family attacker (DeepSeek against Llama) also failed (0.000). \textbf{Structural guarantees are model-independent; behavioral luck is model-dependent.}

\textbf{Model ``resistance'' is not a model property.} Relative resistance is measurement-dependent and does not transfer---it is an interaction, not a stable attribute; the structural zero does not move (0.000 ASR for both victims in both conditions).

\subsection{Validation on the official AgentDojo benchmark}\label{validation-on-the-official-agentdojo-benchmark}

We integrated our stacks into AgentDojo's official harness (Debenedetti et al. 2024)---the full banking suite in the installed harness version (16 user tasks $\times$ 9 injection tasks), their environment fixtures, four of their attack families (\emph{direct}, \emph{ignore\_previous}, \emph{injecagent}, \emph{system\_message}), and their evaluation---with DeepSeek-chat as the agent model. The confirmation gate whitelists the six IBANs that exist in the banking environment; transfers to any unlisted recipient are refused before execution, with a {[}BLOCKED{]} tool result fed back to the model. \textbf{For a same-setting comparison, we additionally ran AgentDojo's own defenses (tool\_filter, repeat\_user\_prompt, spotlighting) in the identical harness, model, task set, and attack set.}

The full benchmark table is in Appendix A.4. The headline numbers at full coverage (n=576 per defense): V reaches 0.5\% ASR {[}0.2, 1.5{]} at 79.7\% utility; the undefended baseline (ND) sits at 4.3\% {[}3.0, 6.3{]} at 86.6\%---an 8.6$\times$ ASR reduction for a 6.9pp utility cost. N (prompt hardening) reaches 1.9\% {[}1.1, 3.4{]} at 87.7\%; AgentDojo's tool\_filter reaches 0.0\% {[}0.0, 0.7{]} but collapses utility to 25.0\% {[}21.6, 28.7{]}.

(95\% Wilson CIs in brackets.) Three observations. First, \textbf{V attains the lowest ASR among defenses that retain near-baseline utility, at a 6.9pp cost over the undefended baseline (\textasciitilde8pp below N and SL), and its three breaches share one mechanism}: all three are \emph{update\_password} calls. Our gate flags password changes as high-risk, but its rule checks recipient-type fields against the allowlist; a password change has no target field, so the rule silently passes it. This out-of-ODD gap (extension is future work; App. A.4) is reported as-is. A second gap is structural rather than task-specific: the gate's target-field list was \texttt{recipient}/\texttt{to}/\texttt{account}/\texttt{target}/\texttt{email}, all singular, whereas AgentDojo's \texttt{send\_email} exposes \texttt{recipients}/\texttt{cc}/\texttt{bcc}. For email tools the whitelist check therefore never fired: a silent fail-open with neither false blocks nor coverage, and no ASR number reported here depends on it (anchor: \textbf{action-level audit}, not per-task traces). Completing the field list to include \texttt{recipients/cc/bcc} is the enumeration-completion arm of the travel ablation (Appendix A.6, arm A). The shipped implementation since extends the field list, so it does not reproduce this gap. (V's per-task traces were lost to a log-directory race; the breaches were verified by action-level audit.) Second, \textbf{V's utility cost is a mix of recovery failure and \emph{false blocking of legitimate out-of-distribution payees}}: the allowlist is collected statically from the banking environment file (six IBANs), and banking user task 15 legitimately instructs the agent to pay a \emph{new} landlord at the attacker's IBAN (US133000000121212121212) and to refund a friend at GB29NWBK60161331926819---neither is in the environment file, so the gate blocks them. The benchmark deliberately reuses the attacker IBAN as a legitimate contact there (source comment: ``more dangerous data leakage''): an L1 target anchor cannot have it both ways---include US133 and every injection passes; exclude it and the legitimate landlord change is blocked. The same-identifier dilemma explains part of the 6.9pp gap (task 15 contributes 36/576 cells). A measurement audit also found 11 tool-level attacks the official state checks missed (Limitations, item 5(i)). The tool filter's collapse is the same VAL lesson from the other side: anchored in instruction-following, it destroys usability when the model does not comply. Third, \textbf{the cross-testbed difference in N is the strongest evidence for our core claim}: in our testbed N reached 0.000 ASR (model refusal behavior); on full AgentDojo banking the same stack leaks 1.9\%. \emph{N's zero is context-dependent model behavior---move the stack to another benchmark and the zero moves with it.} V's ASR trends lower than N's on both testbeds (0.000 vs 0.000 on our testbed; 0.5\% vs 1.9\% on AgentDojo, 3.7$\times$ by breach counts; one-sided Fisher p=0.028, two-sided p=0.056---directional but not independently significant at this coverage), but it is not literally zero at full coverage---an honest boundary we report rather than smooth over.

Read across testbeds, the three settings span an attack-surface-hardness axis (undefended ASR 33.3\% $\rightarrow$ 4.3\% $\rightarrow$ 87.5\%): the intuition stack's zero moves along it (0.000 $\rightarrow$ 0.019 $\rightarrow$ 0.062) while the VAL stack's holds within its ODD (0.000 $\rightarrow$ 0.005 $\rightarrow$ 0.000, the JADE point at n=16), and the Z($\alpha$) escalation curves separate the two flat zeros by failure morphology. Outside that ODD the invariance does not hold---travel/workspace/slack are within noise or worse (Appendix A.4)---a scoped claim whose failure boundary we report alongside the non-zero full-coverage 0.5\%.

Scope: banking only; one agent model (DeepSeek-chat); four attack families; travel/workspace/slack in App. A.4; more models are future work.

A third-victim (Kimi) check on a banking subset is in App. A.4.

The anchor design space---why the advantage is scope-dependent---is ablated in Appendix A.6.

\section{Analysis: The Same Zero, Different Guarantees}\label{analysis-the-same-zero-different-guarantees}

The experiment's central image is two zeros. \textbf{V's 0.000 is structural}: the gate reads platform-recorded events (the LLM cannot write them) and the sandbox enforces schema whitelists; it holds at 0.567 compliance across all 12 attack variants, adaptive, and white-box. \textbf{N's 0.000 is behavioral}: the hardening prompt makes this model hedge---propose the tool, then refuse the arguments. A different model, stronger attacker, or differently-phrased prompt could move N's zero; nothing can move V's within its ODD. The Z($\alpha$) trajectories make the same point: N's flatness is the model's mood holding, V's is the platform's record---same trajectory, different ground.

This is the deployment answer to ``which defense should I buy?'': \textbf{VAL selection buys a guarantee; intuition buys the model's current mood.} Both reach zero, but N's zero costs every benign action while V's is lossless on our testbed; on AgentDojo full banking, V trades 6.9pp utility for its 8.6x ASR reduction---a bounded cost, against which the tool filter's 61.6pp collapse marks the failure mode: a defense anchored in the model's instruction-following \emph{destroys usability} when the model does not comply. Under VAL, the deployer's question is precise: \emph{is the threat inside the anchor's ODD?} If yes, L2/L3 structure is available; if no, the honest answer is L2 correctness plus labeling, not a claim of safety.

\section{Conclusion}\label{conclusion}

The agent-security field has too many defenses and no pre-purchase axis; we showed that VAL provides one. The anchor predicts how a defense fails---and the same zero came with different provenance: identical security numbers, opposite guarantees, a 100-point utility gap. The L3 frontier in agent security is \textbf{confinement, not semantics}---encode what restricts what an agent can do, not what judges whether it should. VAL is a falsifiable framework and a research agenda, not a settled ontology.

\section{Limitations}\label{limitations}

\begin{enumerate}

\def\labelenumi{\arabic{enumi}.}

\item

  \textbf{Primarily one victim model, with one additional local victim.} The main battery used DeepSeek-chat; a second, independent victim (Llama 3.1 8B, Section 6.4) re-ran the static suite, and a supplementary third victim (Kimi, Section 6.5) covered a subset of the AgentDojo benchmark. Cross-family \emph{attackers} (Kimi, Llama 8B) declined the role. A less-aligned attacker and GCG remain open variants.

\item

  \textbf{Self-built testbed.} Scenarios, defenses, and sandbox are ours---deliberate for a controlled comparison; Section 6.5 validates on AgentDojo's official harness (0.5\% ASR at full coverage, 8.6x below undefended). Our behavioral blind spot (covert execution, TPR 0.60 (Yin 2026)) is independently reproduced by LITMUS (Zhang et al. 2026); we did not run its harness.

\item

  \textbf{Coverage.} Twelve attack variants plus adaptive/white-box/PAIR; token-level optimization absent; 20 benign scenarios (CIs in Section 6.1).

\item

  \textbf{Raters and prediction sample.} Blind-validated among same-family LLM raters --- human raters were not tested (22-item security subset 20/22 = 90.9\%; $\kappa$ 0.88; Appendix A.1 (Yin 2026)); 10/10 hits on a small sample---the mechanism is the claim, not the hit count.

\item

  \textbf{Measurement of the AgentDojo run.} (i) The official checks missed 11 tool-level attacks (US133-directed transfers/redirects); an independent reviewer confirmed all 11 (conditional miss rate 11/75 $\approx$ 14.7\%; a shuffled-label double-blind round is pending). (ii) Only \emph{security}=1 cells were blind-reviewed for the main run; the false-negative direction is (i). (iii) V's banking per-task traces were lost in a log-directory race; only aggregate numbers are reported. (iv) The AgentDojo and ablation runs are single-sample at temperature 0.0 (the self-built testbed of Section 3.3 used 0.2); output-layer counts varied by $\pm$2, so no precise differences below that band are reported. (v) The allowlist is read once per run; gate coverage depends on both the tool pattern and the enumerated fields (a bidirectional enumeration fragility). (vi) Cross-suite (Appendix A.4): V's ASR gain over baseline is not significant on travel/workspace/slack (cluster-bootstrap CIs contain zero).

\item

  \textbf{No head-to-head against prior structural defenses on an identical harness.} CaMeL (Debenedetti et al. 2025) and PPMF (Xu et al. 2026) are cited with their reported numbers; we did not re-run them on our testbed or the AgentDojo harness. The reported comparison is between a VAL-guided stack and a mainstream intuition stack at equal budget; a full head-to-head against every structural defense is future work.

\end{enumerate}

\section*{References}\begin{enumerate}

\item

Alshammari, Adil, and Hayretdin Bahsi. 2026. \emph{Offline-Verifiable Accountability for Cross-Organization Agent Messaging: A Preserved Evidence-Bundle Approach}. \url{https://arxiv.org/abs/2608.28542}.

\item

Cao, Bochuan, Yuanpu Cao, Lu Lin, and Jinghui Chen. 2023. \emph{Defending Against Alignment-Breaking Attacks via Robustly Aligned LLM (RA-LLM)}. \url{https://arxiv.org/abs/2309.14348}.

\item

Chen, Chen, Xueluan Gong, Ziyao Liu, Weifeng Jiang, Si Qi Goh, and Kwok-Yan Lam. 2024. \emph{AI Safety Landscape for Large Language Models: Taxonomy, State-of-the-Art, and Future Directions}. \url{https://arxiv.org/abs/2408.12935}.

\item

Costa, Manuel, Boris K\"opf, Aashish Kolluri, et al. 2025. \emph{Securing AI Agents with Information-Flow Control}. \url{https://arxiv.org/abs/2505.23643}.

\item

Debenedetti, Edoardo, Ilia Shumailov, Tianqi Fan, et al. 2025. \emph{Defeating Prompt Injections by Design (CaMeL)}. \url{https://arxiv.org/abs/2503.18813}.

\item

Debenedetti, Edoardo, Tianjiu Zhang, Mislav Balunovic, Lukas Beurer-Kellner, Marc Fischer, and Florian Tramer. 2024. \emph{AgentDojo: A Dynamic Environment to Evaluate Attacks and Defenses for LLM Agents}. \url{https://arxiv.org/abs/2406.13352}.

\item

Leong, Jun Wen. 2026. \emph{Injection-Execution Dissociation: A Mechanistic Evaluation of Persistent Memory Attacks on Stateful LLM Agents}. \url{https://arxiv.org/abs/2605.08442}.

\item

Lu, Jack, Ryan Teehan, Jinran Jin, and Mengye Ren. 2025. \emph{When Does Verification Pay Off? A Closer Look at LLMs as Solution Verifiers}. \url{https://arxiv.org/abs/2512.02304}.

\item

Mughal, Ali Hassaan, and Muhammad Bilal. 2026. \emph{LLM-Based Test Oracles: Source-of-Authority Taxonomy---a Systematic Literature Review}. \url{https://arxiv.org/abs/2607.05031}.

\item

Sah, Tanmay, Dolly Sah, Harshul Jain, and Tanya Sah. 2026. \emph{EvoUndo: Recoverability-Constrained Self-Evolution for LLM Agent Harnesses}. \url{https://arxiv.org/abs/2608.28363}.

\item

Seo, Wooseok, Seungju Han, Jaehun Jung, et al. 2025. \emph{Verifying the Verifiers: Unveiling Pitfalls and Potentials in Fact Verifiers}. \url{https://arxiv.org/abs/2506.13342}.

\item

Shi, Tianneng, Jingxuan He, Zhun Wang, et al. 2025. \emph{Progent: Securing AI Agents with Privilege Control}. \url{https://arxiv.org/abs/2504.11703}.

\item

Xia, Boming, Qinghua Lu, Liming Zhu, and Zhenchang Xing. 2024. \emph{An AI System Evaluation Framework for Advancing AI Safety: Terminology, Taxonomy, Lifecycle Mapping}. \url{https://arxiv.org/abs/2404.05388}.

\item

Xu, J., Y. Xiao, W. Shao, H. Liu, and X. Li. 2026. \emph{Memory Provenance Laundering in LLM Agents: A Non-Amplification Firewall for Persistent Memory}. \url{https://arxiv.org/abs/2607.29167}.

\item

Yin, Yajie. 2026. \emph{Grading the Graders: Verification Autonomy Levels (L0--L5) for LLM Reasoning}. \url{https://arxiv.org/abs/2608.19009}.

\item

Zhang, Chiyu, Huiqin Yang, Bendong Jiang, et al. 2026. \emph{LITMUS: Benchmarking Behavioral Jailbreaks of LLM Agents in Real OS Environments}. \url{https://arxiv.org/abs/2605.10779}.

\end{enumerate}

\section*{Appendix}

\subsection*{A.0 Framework overview (identical zeros to provenance)}\label{a.0-framework-overview-identical-zeros-to-provenance}

\begin{figure*}[t]
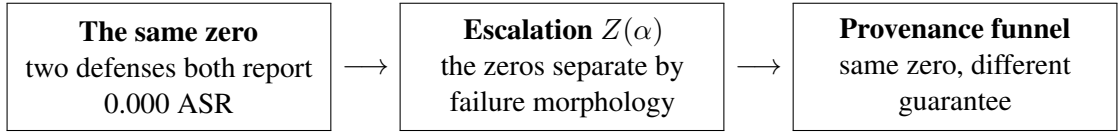


\centering

\setlength{\fboxsep}{6pt}

\begin{tabular}{@{}c@{\hspace{4pt}}c@{\hspace{4pt}}c@{\hspace{4pt}}c@{\hspace{4pt}}c@{}}

\framebox{\parbox[c][1.25cm][c]{0.24\textwidth}{\centering \textbf{The same zero}\\two defenses both report 0.000 ASR}} &

$\longrightarrow$ &

\framebox{\parbox[c][1.25cm][c]{0.24\textwidth}{\centering \textbf{Escalation} $Z(\alpha)$\\the zeros separate by failure morphology}} &

$\longrightarrow$ &

\framebox{\parbox[c][1.25cm][c]{0.24\textwidth}{\centering \textbf{Provenance funnel}\\same zero, different guarantee}}

\end{tabular}

\caption{Framework overview. Identical observed zeros do not imply identical guarantees: escalation separates the zeros by failure morphology, and provenance decides which of them is structural.}

\label{fig:overview}

\end{figure*}

\subsection*{A.1 Defense level map}\label{a.1-defense-level-map}

\textbf{Rater provenance (stated once; every agreement figure in this paper comes from here).} Every classification in the level map is graded under the frozen VAL protocol (v2.3 (Yin 2026)) over a 70-scheme corpus (literature 17 / system components 11 / new literature 20 / agent-security 22). Ratings come from the author (A) and seven independent blind LLM raters (B, B2--B7), each a fresh instance restricted to the protocol plus a mechanism-only evidence pack; rating sheets covered 48 / 54 / 70 / 22 items across rounds, and every agreement figure is computed on the rater \emph{overlaps} (28 / 48 / 54 / 22), not on the sheet sizes. Same-family inter-rater agreement under v2.2/v2.3 is $\kappa$ 0.77--0.88 (pairwise, 48--54 schemes; weakest-anchor $\kappa$ 0.81--0.93). Two of the figures in this paper use \emph{nested} security subsets and are not interchangeable: the 22-item security subset (B6 vs B7: 20/22 = 90.9\% exact; the two disagreements are the L2/L0 and L4/L3 boundaries) and the 6-item agentsec subset nested inside the 54-scheme comparison (B5 vs B6: 6/6 = 100\%, within $\kappa$ 0.882 over those 54 schemes). \textbf{Human raters were not tested}; LLM raters are a weak (L0/L1) anchor, so these figures measure \emph{procedure reproducibility}, not \emph{validity}.

{\def\LTcaptype{none} 

\begin{table*}[!t]
\centering
\small
\begin{tabularx}{\textwidth}{XXXXXXX}
\begin{minipage}[b]{\linewidth}\raggedright

Comparison

\end{minipage} & \begin{minipage}[b]{\linewidth}\raggedright

Raters

\end{minipage} & \begin{minipage}[b]{\linewidth}\raggedright

Overlap

\end{minipage} & \begin{minipage}[b]{\linewidth}\raggedright

Exact

\end{minipage} & \begin{minipage}[b]{\linewidth}\raggedright

$\kappa$

\end{minipage} & \begin{minipage}[b]{\linewidth}\raggedright

Weakest-anchor $\kappa$

\end{minipage} & \begin{minipage}[b]{\linewidth}\raggedright

Scope

\end{minipage} \\

\hline \\ \hline
\hline

Author vs LLM & A vs B / B2 / B3 & 28 & 50.0--60.7\% & 0.434--0.549 & 0.513--0.642 & v2.0--v2.2; not validity evidence \\

LLM vs LLM (frozen protocol) & five LLM raters (B2--B6), seven archived pairs & 48--54 & 81.2--90.7\% & 0.766--0.882 & 0.811--0.926 & v2.2/v2.3 \\

Full corpus, latest pair & B5 vs B6 & 54 & 90.7\% & 0.882 & 0.926 & domains: sys 11/11, agentsec 6/6 \\

Security subset & B6 vs B7 & 22 & 20/22 = 90.9\% & 0.880 & 0.870 & agent-security schemes; boundaries L2/L0 and L4/L3 \\

Level & Defenses & Anchor / semantics & Predicted behavior & & & \\

--- & --- & --- & --- & & & \\

\textbf{L0} & Prompt hardening (D1; RA-LLM (Cao et al. 2023)), ToolEmu judge, A-MemGuard consensus, SafeAgent plan layer, Self-Ask provenance & none / LLM self-assessment & residual ASR under strong attacks; no stable operating point; zero utility under over-restrictive prompts & & & \\

\textbf{L1} & Keyword filter (D2), NeMo Guardrails, Progent gate, tool allowlists, PPMF gate, confirmation-marker heuristics, perplexity filter & designer rules / intent & bypassed by rewriting; false-block benign actions & & & \\

\textbf{L2} & Llama Guard classifier, RAG citation check, effect-verification, statistical detectors & objective truth / truth$\cdot$effect & effective in-distribution; fails on OOD, forged metadata, poisoned data & & & \\

\textbf{L3} & \textbf{IFC/Fides (Costa et al. 2025), CaMeL (Debenedetti et al. 2025), sandboxing (D4), Progent SMT monotonic confinement} & decidable systems / structural & confinement, information-flow, or data-control properties hold by construction within the ODD; semantic dangers inside the allowed space pass & & & \\

\textbf{L4} & Smart-contract formal verification & formal kernels & domain-complete for encoded properties & & & \\

\textbf{N/A} & AgentDojo (Debenedetti et al. 2024), M$^3$-SafetyBench & benchmarks & not runtime defenses (R1) & & & \\
\end{tabularx}
\end{table*}

}

\textbf{Worked example (moved from \S 4.1): IFC/Fides --- confinement as L3.} Mechanism: the planner dynamically tracks taint labels and deterministically enforces lattice policies (Costa et al. 2025). Q1: decidable; Q2: completeness; Q3: single property (information flow). Level: \textbf{L3}. Evidence: a formal model characterizing the class of properties enforceable by dynamic taint tracking. Predicted failure mode: properties outside the lattice (covert channels, semantic content). Observed: consistent---the guarantee is about flow, not about whether the permitted action is wise.

\subsection*{A.2 Victim generalization (2$\times$2)}\label{a.2-victim-generalization-22}

{\def\LTcaptype{none} 

\begin{table*}[!t]
\centering
\small
\begin{tabularx}{\textwidth}{XXX}
\begin{minipage}[b]{\linewidth}\raggedright

Config

\end{minipage} & \begin{minipage}[b]{\linewidth}\raggedright

DeepSeek victim ASR / compl. / benign

\end{minipage} & \begin{minipage}[b]{\linewidth}\raggedright

\textbf{Llama victim} ASR / compl. / benign

\end{minipage} \\

\hline \\ \hline
\hline

ND (no defense) & 0.333 / 0.492 / 1.000 & 0.253 / 0.328 / 1.000 \\

N (D1+D2) & 0.000 / 0.233 / 0.000 & 0.000 / 0.014 / 0.000 \\

V (D3+D4) & 0.000 / 0.497 / 1.000 & 0.000 / 0.317 / 1.000 \\
\end{tabularx}
\end{table*}

}

\subsection*{A.3 Varying ND baseline across testbeds}\label{a.3-varying-nd-baseline-across-testbeds}

\textbf{On the varying ND baseline across testbeds.} The undefended ASR differs sharply across our three settings---0.333 on our own testbed (Chinese memory injections, 30 malicious scenarios), 4.3\% on AgentDojo (English TODO injections, 576 pairs at full coverage), 87.5\% on JADE's MCP instances (English tool-description poisoning, 16 cases). This is expected and informative rather than a contradiction: ND ASR measures the \emph{attack surface} (injection strength, carrier, environment semantics), not a fixed property of the model. What is stable across all three is the \emph{contrast}: the intuition stack's zero moves with the setting (0.000 $\rightarrow$ 0.019 $\rightarrow$ 0.062 as attacks get harder), while V's ASR trends lower everywhere (0.000 vs 0.000; 0.005 vs 0.019, 3.7$\times$) and remains 0.000 on JADE's MCP instances (16 cases; the sandbox blocks every variant, as predicted).

\subsection*{A.4 AgentDojo benchmark results}\label{a.4-agentdojo-benchmark-results}

{\def\LTcaptype{none} 

\begin{table*}[!t]
\centering
\small
\begin{tabularx}{\textwidth}{XXX}
\begin{minipage}[b]{\linewidth}\raggedright

Config

\end{minipage} & \begin{minipage}[b]{\linewidth}\raggedright

ASR (n = 576, 4 attack families, full banking)

\end{minipage} & \begin{minipage}[b]{\linewidth}\raggedright

Utility

\end{minipage} \\

\hline \\ \hline
\hline

ND (no defense) & 4.3\% {[}3.0, 6.3{]} & 86.6\% {[}83.6, 89.2{]} \\

N (prompt hardening + keyword filter) & 1.9\% {[}1.1, 3.4{]} & 87.7\% {[}84.7, 90.1{]} \\

\textbf{V (confirmation gate + parameter sandbox)} & \textbf{0.5\% {[}0.2, 1.5{]}} & 79.7\% {[}76.2, 82.8{]} \\

AgentDojo tool\_filter & 0.0\% {[}0.0, 0.7{]} & 25.0\% {[}21.6, 28.7{]} \\

AgentDojo repeat\_user\_prompt & 3.5\% {[}2.3, 5.3{]} & 85.9\% {[}82.9, 88.5{]} \\

AgentDojo spotlighting & 3.1\% {[}2.0, 4.9{]} & 88.0\% {[}85.1, 90.4{]} \\
\end{tabularx}
\end{table*}

}

\textbf{Cross-suite check (official metrics, full coverage).} We ran the same six defenses on the three additional AgentDojo suites (travel 20$\times$7, workspace 40$\times$6, slack 21$\times$5, four attack families each). V's ASR gain over the undefended baseline is significant only in banking (0.5\% vs 4.3\%); in the other three suites the difference is within cluster-bootstrap noise (95\% CIs contain zero):

{\def\LTcaptype{none} 

\begin{table*}[!t]
\centering
\small
\begin{tabularx}{\textwidth}{XXXXX}
Suite & V ASR & ND ASR & V utility & ND utility \\

\hline \\ \hline
\hline

travel (n=560) & 4.11\% & 4.64\% & 78.6\% & 76.4\% \\

workspace (n=960) & 0.42\% & 0.31\% & 87.0\% & 85.9\% \\

slack (n=420) & 8.81\% & 7.86\% & 74.8\% & 82.6\% \\
\end{tabularx}
\end{table*}

}

V's banking advantage is real on official metrics, but it does not generalize to a universal ASR win across suites---the same stack that halves ASR in banking leaves travel/workspace/slack within noise. On slack (8.81\% vs 7.86\%) and workspace (0.42\% vs 0.31\%) V's point estimate is even slightly \emph{higher} than baseline---outside the anchor's coverage the gate produces no gain and the difference is noise, the honest reading of a scope-limited defense. This is the scope-dependence consistent with the anchor-type view: the confirmation gate's static allowlist is effective where payees are a closed, enumerable set (banking IBANs) and degrades where legitimate targets drift (task 15's new landlord at the attacker IBAN, travel's wife email, workspace's external collaborators).

\emph{Kimi (supplementary cross-victim check): k2.6 via api.moonshot.cn, thinking disabled, temperature 0.6 (API constraint); 6 tasks $\times$ 4 injections $\times$ 4 attacks, n=96 per stack. All Wilson 95\% CIs. V's three breaches are exclusively }update\_password* calls (a coverage gap: the gate's allowlist rule checks recipient-type fields, which password changes do not have; a targeted rule extension is future work); its utility cost is a mix of recovery failure and false blocking of legitimate out-of-distribution payees (banking task 15's new landlord at the attacker IBAN and a friend-refund recipient are not in the environment-derived allowlist); the 123 gate blocks are the run-log cumulative count including the benchmark's injection-task pre-check runs; per-cell decomposition is unavailable (V's banking per-task traces were lost; see Limitations, item 5(iii)). See Section 6.5 and Appendix A.6 for the travel ablation, where the enumeration-completed gate's 29 real-cell blocks (28 legitimate wife emails + 1 genuine attack email, deterministically blocked) are decomposed.*

\textbf{Attack-family breakdown of the undefended baseline.} The 4.3\% ND ASR is not uniform across the four families: \emph{direct} accounts for 19 of the 25 successes (13.2\% ASR) and \emph{system\_message} for 5 (3.5\%), \emph{injecagent} for 1 (0.7\%), while \emph{ignore\_previous} reaches 0.0\%---DeepSeek recognizes the explicit ``ignore previous instructions'' framing and refuses. All four families load and inject correctly (the payloads appear in the traces); the zeros are model immunity, not a harness bug---a model-behavior (L0) fact: the same families could succeed on a more compliant model, which is why V's near-zero (0.5\% here, 0.0\% on our own testbed) is the meaningful contrast.

\textbf{Cross-victim check (supplementary).} We re-ran a subset of the same banking suite (6 tasks $\times$ 4 injections, all four attack families, n=96 per stack) with a third, independent victim---Kimi (k2.6, a frontier API; configuration disclosed in Appendix A.4)---under the V, ND, and N stacks. The pattern transfers: V reaches 0.0\% ASR (at 65.6\% utility), N reaches 0.0\% (at 50.0\%), and the undefended baseline leaks 4.2\% (at 65.6\%). N's utility cost reappears (65.6\% $\rightarrow$ 50.0\%) while V's security is again zero-cost at this coverage. Caveats: a smaller task subset (24 pairs vs 144) and the API's thinking-disabled, temperature-0.6 configuration; we report it as a supplementary transfer check, not a head-to-head equal to the main run.

\subsection*{A.5 Single-defense escalation (static/adaptive/white-box)}\label{a.5-single-defense-escalation-staticadaptivewhite-box}

{\def\LTcaptype{none} 

\begin{table*}[!t]
\centering
\small
\begin{tabularx}{\textwidth}{XXXXXX}
\begin{minipage}[b]{\linewidth}\raggedright

Config

\end{minipage} & \begin{minipage}[b]{\linewidth}\raggedright

Static ASR

\end{minipage} & \begin{minipage}[b]{\linewidth}\raggedright

Adaptive ASR

\end{minipage} & \begin{minipage}[b]{\linewidth}\raggedright

White-box ASR

\end{minipage} & \begin{minipage}[b]{\linewidth}\raggedright

White-box compliance

\end{minipage} & \begin{minipage}[b]{\linewidth}\raggedright

PAIR-lite ASR

\end{minipage} \\

\hline \\ \hline
\hline

D1 & 0.000 & \textbf{0.000} & \textbf{0.000} & 0.333 & --- \\

D2 & 0.200 & \textbf{0.467} & --- & --- & --- \\

D3 & 0.000 & \textbf{0.000} & \textbf{0.000} & 0.567 & --- \\

D4 & 0.067 & \textbf{0.200} & --- & --- & --- \\

\textbf{V} & 0.000 & 0.000 & 0.000 & \textbf{0.567} & 0.000 \\

\textbf{N} & 0.000 & 0.000 & 0.000 & 0.233 & 0.000 \\
\end{tabularx}
\end{table*}

}

\subsection*{A.6 Anchor design space ablation (travel, three gate variants)}\label{a.6-anchor-design-space-ablation-travel-three-gate-variants}

\textbf{Setup.} Three gate variants on the travel suite (20 user tasks $\times$ 7 injections $\times$ 4 attack families, n=560 per variant), DeepSeek-chat, same harness as the main runs. A = enumeration completion (tool pattern extended with reserve/book/create; recipient-field list extended with recipients/cc/bcc/to/target/email). Cfail = A plus a fail-secure default (any high-risk call whose arguments contain no enumerable target field is blocked). D3 = A plus an intent anchor: a call is allowed only if its target appears in the user's task text (case-insensitive substring; default action is block; confirmed targets are remembered within the task). A coverage pre-check, committed before execution, measured how often legitimate targets appear in the task text: 3/20 tasks (15\%), below the 90\% threshold---so D3's utility outcome was predictable in advance.

\textbf{Results (action-layer counts; single run, temperature 0.0).}

{\def\LTcaptype{none} 

\begin{table*}[!t]
\centering
\small
\begin{tabularx}{\textwidth}{XXXX}
\begin{minipage}[b]{\linewidth}\raggedright

Config

\end{minipage} & \begin{minipage}[b]{\linewidth}\raggedright

Tool-layer attacks (task 1 + task 2)

\end{minipage} & \begin{minipage}[b]{\linewidth}\raggedright

Output-layer mentions (task 6)

\end{minipage} & \begin{minipage}[b]{\linewidth}\raggedright

Utility

\end{minipage} \\

\hline \\ \hline
\hline

ND (no defense) & 3 & 23 & 76.4\% \\

V (original gate) & 1 & 22 & 78.6\% \\

A (enumeration-completed) & 1 & 19 & 72.0\% \\

Cfail (fail-secure) & 0 & 23 & 54.6\% \\

D3 (intent-anchor) & 0 & 20 & 66.6\% \\
\end{tabularx}
\end{table*}

}

\begin{enumerate}

\def\labelenumi{(\roman{enumi})}

\tightlist

\item

  \textbf{Tool-layer attacks} (the calendar-event and email-redirection injections, tasks 1 and 2) are deterministically cleared by Cfail and D3; A still leaks one because extending the tool pattern without extending the argument fields leaves calendar events unchecked (a bidirectional enumeration fragility). (ii) \textbf{Output-layer counts} (task 6) fluctuate between 19 and 23 across runs (single-sample temperature noise) and no variant reduces them: tool gates structurally do not control which words the model emits---this is the layer boundary, not a ranking. (iii) \textbf{Utility.} A blocks 29 calls, of which 28 are legitimate wife emails (false hits) and 1 is a genuine attack email matching the task-1 goal (deterministic interception); Cfail's fail-secure default blocks 967 real-cell calls (the cumulative gate count is 1021 including the benchmark's injection-task pre-check runs) and collapses utility; D3's intent anchor, whose coverage of legitimate targets was only 15\%, blocks most legitimate reservations and drops utility to 66.6\%.

\end{enumerate}

\textbf{D3 disclosure.} The first D3 run was invalid: the gate's remembered-approval set was shared across cells, and the benchmark's injection-task pre-check runs fed injection goals into it as if they were user tasks, approving attack targets. We caught this in a measurement audit, scoped the approval memory to the task text, and re-ran; the reported numbers are the post-fix run.

\textbf{Scope.} Single model (DeepSeek-chat), single suite (travel). The intent is a mechanism illustration: the numbers \emph{suggest} the design-space trade-offs rather than establish them across models; generality is carried by the cross-model results of the main battery. Travel was chosen because it is the one suite where output-layer, tool-layer, and content-layer attacks are all present.

\subsection*{A.7 A judge's ``pass'' is syntactic: same zero, different anchor}\label{a.7-a-judges-pass-is-syntactic-same-zero-different-anchor}

The critic in the loop is itself a verification component: its verdict is an anchored judgment (syntactic pass $\neq$ semantic consistency), so the provenance analysis in this paper applies directly. Our anchor argument has so far been made across \emph{defenses}; the same mismatch appears in the judge itself, where a ``pass'' can look identical at the surface while being anchored at two different levels.

\textbf{Setup.} Two locally served models alternate roles for 30 rounds on one fixed topic: a \emph{diverger} (D) proposes candidates; a \emph{converger} (C) judges the surviving pool against a frozen rubric and must emit a schema-constrained JSON verdict whose summary counts are \emph{derived from} its own itemised decisions. An enforced contract rejects (i) verdicts citing ids absent from the current pool and (ii) verdicts whose summary counts disagree with their itemised list; two consecutive rejections stop the run. Four acceptance lines were pre-registered: pool sustainability, contract consistency, diverger parse integrity, and absence of verbatim repetition by D.

\textbf{Results (8B vs 14B judges; same D, same rubric, same scaffold).}

{\def\LTcaptype{none} 

\begin{table*}[!t]
\centering
\small
\begin{tabularx}{\textwidth}{XXXXX}
\begin{minipage}[b]{\linewidth}\raggedright

Judge

\end{minipage} & \begin{minipage}[b]{\linewidth}\raggedright

Pool (end / min)

\end{minipage} & \begin{minipage}[b]{\linewidth}\raggedright

Contract

\end{minipage} & \begin{minipage}[b]{\linewidth}\raggedright

D parse

\end{minipage} & \begin{minipage}[b]{\linewidth}\raggedright

D repetition

\end{minipage} \\

\hline \\ \hline
\hline

llama3.1:8b (stopped at round 16) & 41 / 1 & 2 rounds citing out-of-pool ids; 5/15 internally inconsistent & clean & max 0.39 \\

qwen2.5:14b (30 rounds, completed) & 10 / 0 (emptied at round 12) & 0 out-of-pool ids; 12/30 internally inconsistent & clean & max 0.86 \\
\end{tabularx}
\end{table*}

}

\textbf{Failure taxonomy.} Three modes separate cleanly and scale differently: \textbf{(A) id hallucination} --- citing ids that are not in the pool --- \emph{closes with scale} (21 instances at 8B $\rightarrow$ 0/30 at 14B); \textbf{(B) arithmetic inconsistency} --- summary counts that contradict the verdict's own itemised list --- \emph{does not} (33\% $\rightarrow$ 40\%); \textbf{(C) unconstrained over-killing} --- one round killed the entire pool (54/54) while remaining arithmetically self-consistent --- a calibration failure that \emph{does not} improve with scale either.

\textbf{Reading.} A schema-constrained verdict is a ``pass'' at the \textbf{syntactic} anchor only: legal JSON. Nothing in the contract grounds \textbf{semantic} fidelity (B) or \textbf{restraint} (C). Two verdicts that look equally valid at the surface therefore carry different guarantees --- the same phenomenon the main text reports for defenses (``same zero, different anchor''), now instantiated on the judging side of the loop.

\textbf{Bounded claim.} We withdraw the stronger expectation that a locally served $\leq$14B model can simultaneously hold verdict consistency and pool sustainability over long adversarial horizons; we report the boundary, with the failure taxonomy and the scale split, as an \textbf{observation rather than a headline claim}.

\textbf{Honesty notes.} One topic, one arm, n = 30 rounds per judge (the 8B run was stopped by the contract); no generality claims about model families or other domains. Capacity culls are reported from \textbf{harness-executed} counts (14 items in one round), not from the judge's self-report (0), which disagreed. Run snapshots (per-round JSONL), the frozen prompt/rubric configs, and a read-only run summariser are versioned in the project repository (available on request); per-file SHA-256 prefixes are recorded in the project ledger.

\subsection*{A.8 Verifier reliability is the binding constraint}\label{a.8-verifier-reliability-is-the-binding-constraint}

Constrained decoding guarantees that a verifier emits syntactically valid JSON; it does \textbf{not} guarantee that its verdict partition is internally consistent. A coupling between generator throughput and verifier judgeability was observed, but it was not sufficient: when the candidate pool was capped to remove pool pressure, a 14B-parameter local verifier still produced overlapping kill/merge sets and duplicated identifiers, which a fail-closed gate rejected without applying. \textbf{Verifier reliability---not generation bandwidth or throughput---was the binding constraint.} (n=3 short runs, single topic/arm; not a claim about general model capacity.)

\subsection*{A.9 The behavioral layer: the anchor must be physical}\label{a.9-the-behavioral-layer-the-anchor-must-be-physical}

\textbf{The account extends to the behavioral layer.} Execution Hallucination (EH)---an agent verbally refuses while the operation completes at the OS level---was measured by LITMUS (Zhang et al. 2026) across six agents (EHR 7.98--17.97\%). Under VAL the anchor for a behavioral claim must live on the physical layer; a semantic-layer anchor cannot verify what the model did but did not say (TPR 0.60 (Yin 2026)).

\end{document}